\documentclass[final,5p,times,twocolumn,compress]{elsarticle}

\usepackage{amssymb}
\usepackage{amsmath}
\usepackage{graphicx}
\usepackage{float}
\usepackage{dcolumn}
\usepackage{multirow}
\usepackage{changes}
\usepackage[utf8]{inputenc}
\usepackage[T1]{fontenc}
\usepackage[colorlinks = true,linkcolor = blue,urlcolor  = blue,citecolor = blue,anchorcolor = blue]{hyperref}
\usepackage{enumerate}
\usepackage{times}
\usepackage{microtype}
\usepackage{orcidlink}
\usepackage[percent]{overpic}
\usepackage{stfloats}

\journal{Physics Letters B}

\begin{document}

\begin{frontmatter}


\title{How the HESS J1731-347 object could be explained using $\bf{K^{-}}$ condensation}

\author[first]{M. Veselsk\'y\orcidlink{0000-0002-7803-0109}}
\ead{Martin.Veselsky@cvut.cz}
\author[second,third]{P.S. Koliogiannis\corref{cor1}\orcidlink{0000-0001-9326-7481}}
\ead{pkoliogi@phy.hr}
\author[first]{V. Petousis\orcidlink{0000-0002-5575-6476}}
\ead{vlasios.petousis@cvut.cz}
\author[fourth]{J. Leja\orcidlink{0000-0003-1798-116X}}
\ead{jozef.leja@stuba.sk}
\author[third]{Ch.C. Moustakidis\orcidlink{0000-0003-3380-5131}}
\ead{moustaki@auth.gr}

\cortext[cor1]{P.S. Koliogiannis}

\affiliation[first]{organization={Institute of Experimental and Applied Physics, Czech Technical University},
            city={Prague},
            postcode={11000}, 
            country={Czechia}}

\affiliation[second]{organization={Department of Physics, Faculty of Science, University of Zagreb, Bijeni\v cka cesta 32},
            city={Zagreb},
            postcode={10000}, 
            country={Croatia}}
            
\affiliation[third]{organization={Department of Theoretical Physics, Aristotle University of Thessaloniki},
            city={Thessaloniki},
            postcode={54124}, 
            country={Greece}}

\affiliation[fourth]{organization={Faculty of Mechanical Engineering - Slovak University of Technology in Bratislava},
            city={Bratislava},
            postcode={81231}, 
            country={Slovakia}}

\begin{abstract}
The recent observation of a compact star with a mass of $M=0.77^{+0.20}_{-0.17}~{\rm M_{\odot}}$ and a radius of $R=10.4^{+0.86}_{-0.78}$ km, located within the supernova remnant HESS J1731-347, has substantially reinforced the evidence for the presence of exotic matter in neutron stars core. This finding has markedly enhanced our comprehension of the equation of state for dense nuclear matter. In the present work, we investigate the possible existence of a kaon condensation in hadronic neutron stars by employing and comparing two theoretical frameworks: the Relativistic Mean Field model with first order kaon condensate and the Momentum-Dependent Interaction model complemented by chiral effective theory. To the best of our knowledge, this represents a first alternative attempt aimed to explain the bulk properties of the specific object with the inclusion of a kaon condensation in dense nuclear matter. The application of two different models enriches the research, providing insights from the aspect of different theoretical frameworks that accurately predict the existence of HESS J1731-347. In both cases significant insights are extracted for the parameter space of models, emphasizing to those concerning the nucleon-kaon potential, the threshold density for the appearance of a kaon condensation, as well as the parameter $a_{3}m_{s}$ which is related to the strangeness content of the proton. Concluding, the present research indicates that a more systematic investigation of similar objects could offer valuable constraints on the properties of dense nuclear matter.
\end{abstract}

\begin{keyword}
Neutron stars \sep Equation of state \sep Exotic matter \sep Kaon condensation 



\end{keyword}

\end{frontmatter}

\section{Introduction}
The uncertain behavior of the equation of state (EoS) of neutron star (NS) matter at high densities, arising from the lack of experimental data, permits the exploration of the inner core through various theoretical frameworks. Specifically, the inner core of NS is believed to undergo several phase transitions in ultra-dense matter, including the formation of hyperons, a mixed phase of hyperons and quarks, deconfined quarks, and boson condensates~\cite{Griffin_Snoke_Stringari_1995,Brown-1992,Thorsson-1994,Glendenning-1998,Glendenning-1999,Lim-2014,PhysRevD.102.123007,Menezes-2005}. These potential phases exert a profound influence on the structure of the EoS, particularly in shaping the mass-radius relationship, the maximum mass, and the stability of NSs, as well as determining the density thresholds for such transitions.

At densities near that of normal nuclear matter, the behavior of the EoS is well-constrained both from a theoretical and experimental aspect. In this low density region, the EoS is predominately governed by hadrons (neutrons and protons), with additional contributions arising from leptons, such as electrons and muons. As densities increase, the behavior of NS matter becomes less understood and one plausible scenario is the onset of strangeness, manifesting as a boson condensate, including particles such as pions or kaons~\cite{Brown-1992,Thorsson-1994}. Emphasizing to kaons, the strong interaction of negatively charged kaons with the nuclear medium, may sufficiently reduce their effective mass in a way that a boson may replace electrons as the neutralizing agent in charge neutral matter~\cite{Brown-1992,Thorsson-1994,Glendenning-1998}. The charge neutrality condition in NSs favors a kaon condensation because the neutrons at the top of the Fermi sea decay into protons and electrons. As a result, when the electron chemical potential matches the effective kaon mass, the kaons appear into the zero momentum state for conserving the charge neutrality.

Kaon condensation in NSs has been a subject of significant interest in the study of dense matter. The initial purpose for a kaon condensation was the existence of a mechanism that sufficiently lowers the energy of dense nuclear matter, making it energetically favorable for kaons to form a Bose-Einstein condensate at extreme densities. In this field, Kaplan and Nelson~\cite{Kaplan-1986} suggested that the attraction between nucleons and kaons could lead to a kaon condensation at high densities, resulting to a possible softening on the EoS by reducing its maximum mass. Later on, Brown \textit{et al.}~\cite{Brown-1992} proposed a new mechanism based on the decay of energetic electrons to kaons and on the nuclear symmetry energy without requiring ab initio pion condensation. Specifically, the authors found that the pion condensate was needed only to generate a finite chemical potential, introduced to maintain charge neutrality. As a result, in systems where the chemical potential is inherently large, a negatively charged kaon condensation could occur even without the presence of pions. This scenario is particularly relevant for neutron stars, where $\beta$-equilibrium leads to a high chemical potential, reaching hundreds of MeV in dense NS matter. A few years later, Brown \textit{et al.}~\cite{Brown-1994} expanded upon this idea, linking kaons to potential implications for stellar collapse and the formation of compact ``nuclear stars" or low-mass black holes. Subsequently, Thorsson \textit{et al.}~\cite{Thorsson-1994} investigated the effects of kaon condensation on the bulk properties of neutron stars. More recently, Glendenning and Schaffner-Bielich~\cite{Glendenning-1998} introduced models incorporating kaon condensates within a relativistic mean field (RMF) framework, demonstrating that kaons may give rise to mixed phases in neutron star cores, where hadrons coexist with kaon condensates. 

Additionally, kaon condensate with additional degrees of freedoms, such as hyperons and $\Delta$-resonances, and the effects of strongly magnetized dense matter, have been explored in Refs.~\cite{PhysRevD.103.063004,PhysRevC.105.015807,PhysRevC.107.035807,SEDRAKIAN2023104041}. The presence of kaons has also been investigated within the Skyrme model framework by Adam \textit{et al.}~\cite{PhysRevD.107.074007}. Finally, Sharifi \textit{et al.}~\cite{Sharifi_2021} examined kaon condensation in neutron stars, analyzing its impact on bulk properties in relation to binary NS mergers.

Considerable research has been dedicated to interpreting HESS J1731-347 as either normal NS or a hybrid star. Specifically, its properties have been studied within the RMF model by Kubis \textit{et al.}~\cite{PhysRevC.108.045803} and within the mean-field model with a parity doublet structure by Gao \textit{et al.}~\cite{PhysRevC.109.065807}. Hybrid stars comprising quarks or heavy baryons alongside nucleons have also been explored in Refs.~\cite{LI2023138062,PhysRevC.108.025806,PhysRevD.110.043026,Li_2024}. Additionally, Sagun \textit{et al.}~\cite{Sagun_2023} investigated various scenarios in which HESS J1731-347 could be described as a NS, a hybrid star, a strange star, or a star with an admixture of dark matter.

In the present study, we emphasize on the appearance of negatively charged kaons in the form of neutralizing agents in NS matter as a way to describe the ultralight compact object within the supernova remnant HESS J1731-347 with a mass of $M=0.77^{+0.20}_{-0.17}~{\rm M_{\odot}}$ and a radius of $R=10.4^{+0.86}_{-0.78}$ km. In particular, we employ two theoretical nuclear frameworks for nuclear matter: the Momentum-Dependent Interaction (MDI) model~\cite{Koliogiannis-2020} and the RMF model. Concerning the description of the kaon condensation, two approaches had been taken into account, the first order model for a kaon condensate of Glendenning and Schaffner-Bielich~\cite{Glendenning-1998, Glendenning-1999} and the chiral effective theory of Kaplan and Nelson~\cite{Kaplan-1988} which further developed by Brown \textit{et al.}~\cite{Brown-1992}. The aforementioned frameworks lay out significant differences in the approach for the kaon condensation and in-medium interactions. Specifically, we stress out insights related to the nucleon-kaon potential, the threshold density for the appearance of a kaon condensation, as well as the parameter $a_{3}m_{s}$ which is related to the strangeness content of the proton. The results demonstrate that both theoretical frameworks accurately predict the existence of the ultralight compact object. Moreover, they converge at high values of the kaon potential, yielding results that are independent of both the nuclear model and the specific kaon model employed.
To the best of our knowledge, this represents a first alternative attempt aimed to explain the bulk properties of the specific object with the inclusion of a kaon condensation in dense nuclear matter. 

The paper is structured as follows: Sec.~\ref{sec:Theoretical_Framework} provides the theoretical frameworks for the RMF model and for the MDI model, where both approaches emphasizing the kaons interaction, while Sec.~\ref{sec:Results_and_Implications} displays the results of the present study and discusses their implications. Finally, Sec.~\ref{sec:Remarks} contains the scientific remarks.

\section{Theoretical Framework}
\label{sec:Theoretical_Framework}
\subsection{$\beta$-equilibrated matter}
The study of NS matter requires the existence of a chemical equilibrium for all reactions  to be stable at high densities. In the present study, we consider that the NS matter consists of nucleons (neutrons and protons), negatively charged kaons, electrons and muons, while we neglect the corresponding neutrinos by assuming cold NS matter. In general, both $\beta$-decay and inverse $\beta$-decay would take place as~\cite{Thorsson-1994,Lim-2014}:
\begin{equation}
    n \rightarrow p + e^{-} + \bar{\nu}_{e},\quad p + e^{-} \rightarrow n + \nu_{e}.
\end{equation}
As we assumed cold NS matter in which neutrinos have left the system, the chemical equilibrium is established as:
\begin{equation}
    \mu_{n} - \mu_{p} = \mu_{e},
\end{equation}
where $\mu_{n}$, $\mu_{p}$ and $\mu_{e}$ are the chemical potentials of neutrons, protons and electrons, respectively. Furthermore, when the electron Fermi energy is greater than the muon mass $(\mu_{e} > m_{\mu}c^{2})$, it is energetically favorable for electrons to decay into muons and be in a chemical equilibrium state:
\begin{equation}
    e^{-} \rightarrow \mu^{-} + \bar{\nu}_{\mu} + \nu_{e}, \quad \mu_{\mu} = \mu_{e}.
\end{equation}
If the strong interactions of kaons with nucleons in the medium have sufficiently low the effective mass of negatively charged kaons, kaons replace electrons as neutralizing agents according to the following strangeness processes
\begin{equation}
    n \leftrightarrow p + K^{-}, \quad e^{-} \leftrightarrow K^{-} + \nu_{e}.
    \label{eq:strangeness_process}
\end{equation}
Assuming that the in-medium process described through Eq.~\eqref{eq:strangeness_process} take place fast enough to establish chemical equilibrium, the chemical potentials can be described as
\begin{equation}
    \mu_{n} - \mu_{p} = \mu_{K}, \quad \mu_{e}=\mu_{K}=\mu,
\end{equation}
where $\mu_{K}$ is the chemical potential of the negatively charged kaon condensate. It needs to be noted that while the equilibrium process allows the appearance of antiparticles, in the present work we neglect their possible appearance.
\subsection{Nuclear theoretical framework: RMF model}
In the case of the RMF theory using the extended Dirac-Hartree approximation, the energy density and pressure of neutron matter is given by the following expressions \cite{Serot-1997}:
\begin{align}
{\cal E}_{N}&=\frac{(\hbar c)^3g_{\omega N}^2}{2(m_{\omega}c^2)^2}n_N^2+ \frac{(\hbar c)^3(\frac{g_{\rho N}}{2}){^2}}{2(m_\rho c^2)^2}\rho_I^2 \nonumber\\
 &+ \frac{(m_{\sigma}c^2)^2}{2g_{\sigma N}^2(\hbar c)^3}(M_Nc^2-M_N^*c^2)^2\nonumber\\
 &+ \frac{\kappa}{6g_{\sigma N}^3}(M_Nc^2-M_N^*c^2)^3+\frac{\lambda}{24g_{\sigma N}^4}(M_Nc^2-M_N^*c^2)^4\nonumber\\
 &+\sum_{i=n,p} \frac{\gamma}{(2\pi)^3}\int_0^{k_Fi} 4\pi k^2 \sqrt{(\hbar c k)^2+(m_i^* c^2)^2}dk,
 \label{RMF-E-1}
\end{align}
\begin{align}
{\cal P}_{N}&=\frac{(\hbar c)^3g_{\omega N}^2}{2(m_{\omega}c^2)^2}n_N^2+ \frac{(\hbar c)^3 (\frac{g_{\rho N}}{2}){^2}} {2(m_\rho c^2)^2}\rho_I^2 \nonumber \\
 &-\frac{(m_{\sigma}c^2)^2}{2g_{\sigma N}^2(\hbar c)^3}(M_Nc^2-M_N^*c^2)^2\nonumber\\
 &+ \frac{\kappa}{6g_{\sigma N}^3}(M_Nc^2-M_N^*c^2)^3+\frac{\lambda}{24g_{\sigma N}^4}(M_Nc^2-M_N^*c^2)^4\nonumber\\
 &+\sum_{i=n,p}  \frac{1}{3}\frac{\gamma}{(2\pi)^3}\int_0^{k_Fi} \frac{4\pi k^2}{\sqrt{(\hbar c k)^2+(m_i^* c^2)^2}}dk,
 \label{RMF-P-1}
\end{align}
where ${\cal E}_{N}$ is the energy density, ${\cal P}_{N}$ is the pressure, $g_{\sigma N}$, $g_{\omega N}$ and $g_{\rho N}$ are the couplings of the scalar boson, vector boson, and iso-vector $\rho$-meson to nucleons respectively, $m_{\sigma}$, $m_{\omega}$ and $m_\rho$ are the rest masses of scalar and vector bosons and $\rho$-meson respectively, the term $\rho_I$ involves the difference between the proton and neutron densities (important for finite nuclei), also $\kappa$ and $\lambda$ are the couplings of the cubic and quartic self-interaction of the scalar boson, $M_N$ and $M_N^*$ are the rest mass and the effective mass of the nucleon,
$n_N$ is the nucleonic density, $k_F$ is the Fermi momentum of nucleons at zero temperature and $\gamma$ is the degeneracy, with value $\gamma= 4$ for symmetric nuclear matter and $\gamma= 2$ for neutron matter (used in this investigation). 

The kaon condensate was introduced according to the first order kaon condensate (FOKC) model of Glendenning and Schaffner-Bielich ~\cite{Glendenning-1998, Glendenning-1999}. It specifically considers $K^{-}$ particles which can play similar role as electrons in the NS matter. The kaon potential was fixed by the value at saturation density $\rho_0$ of symmetric nuclear matter:
\begin{equation}
U_{K}(\rho_0) = - g_{\sigma K} \frac{(M_{N}-M_{N}^{*}(\rho_0)) c^{2}}{g_{\sigma N}} - 
(\hbar c)^3 g_{\omega K} g_{\omega N} \frac{\rho_0}{m_{\omega}^{2} c^{4}},
\label{uk0}
\end{equation}
where $g_{\sigma K}$ and $g_{\omega K}$ are couplings of $\sigma$ and $\omega$ mesons to $K^{-}$. Kaon chemical potential at given baryonic density is evaluated as: 
\begin{eqnarray}
\mu_{K}(\rho,x_{p}) &=& m_{K} - g_{\sigma K} \frac{(M_{N}-M_{N}^{*}(\rho_{N},x_{p})) c^{2}}{g_{\sigma N}} \nonumber\\  
&-& (\hbar c)^3 g_{\omega K} g_{\omega N} \frac{\rho_{N}}{m_{\omega}^{2} c^{4}} \nonumber\\  
&-& (\hbar c)^3 g_{\rho K} g_{\rho N} \frac{\rho_N}{m_{\rho}^{2} c^{4}}(1 - 2 x_{p}),
\label{muk}
\end{eqnarray}
where  $g_{\rho K}$ is a coupling of $\rho$ meson to $K^{-}$ and $x_{p}$ is proton fraction. The effective mass $M_{N}^{*}(\rho,x_{p})$ is approximated by parabolic dependence on $x_{p}$ between values for symmetric nuclear matter and pure neutron matter. The value of $\mu_{K}(\rho,x_{p})$ is then used to calculate the conditions of chemical equilibrium of the system: 
\begin{eqnarray}
  \mu_{n} - \mu_{p}  = \mu_{e} = \mu_{K}, \\
  \rho_{p}  = \rho_{e} + \rho_{K},
\label{betaeq}
\end{eqnarray}
which provides also the electron and kaon number density.
The energy density for kaons can be expressed as: 
\begin{equation}
\epsilon_{K} = \mu_{K} \rho_{K}.
\label{ek}
\end{equation}
Similar to Refs.~\cite{Glendenning-1998, Glendenning-1999}, kaon condensate is not considered to contribute to pressure. 

\subsection{Nuclear theoretical framework: MDI model}
\subsubsection{Fixed baryon number state}
The theoretical framework for the inclusion of kaon particles in NSs within the MDI model originates from the model introduced by Kaplan and Nelson complemented by the chiral effective theory~\cite{Kaplan-1988}, where the addition of nuclear interactions is based to Thorsson \textit{et al.}~\cite{Thorsson-1994}. In this framework, considering a fixed baryon number state, the effective energy density of the NS matter has the form~\cite{Thorsson-1994}:
\begin{align}
    \tilde{{\cal E}}(u,x,\mu,\theta) &= {\cal E}_{\rm MDI}(u,x=1/2) + u\rho_{0} (1-2x)^{2} S(u) \nonumber \\ 
    &- f^{2}\frac{\mu^{2}}{2 (\hbar c)^{3}}\sin^{2}\theta+f^{2}\frac{2m_K^{2} c^{4}}{(\hbar c)^{3}}\sin^2\frac{\theta}{2} \nonumber \\
    &+ \mu u\rho_{0}x - \mu u\rho_{0}(1+x)\sin^{2}\frac{\theta}{2} \nonumber \\
    &+ (2a_{1}m_{s}x+\mathcal{T}_{23})u \rho_{0}\sin^{2}\frac{\theta}{2} \nonumber\\
    &+ \tilde{{\cal E}}_{e}+\eta(|\mu|-m_{\mu}c^{2})\tilde{{\cal E}}_{\mu},
    \label{eq:eff_energy_density_mdi}
\end{align}
where
$\rho_{0}$ is the nuclear saturation density, $u=\rho/\rho_{0}$ is the nucleon density ratio, ${\cal E}_{\rm MDI}(u,x=1/2)$ is the energy density of symmetric nuclear matter defined through the MDI model~\cite{Koliogiannis-2020}, $S(u)$ is the nuclear symmetry energy, $x$ is the proton fraction, $\mu$ is the chemical potential, $\theta$ is an amplitude parameter for the appearance of kaons, $f=93~{\rm MeV}$ is the pion decay constant, $m_{K}$ is the mass of the negatively charged kaon, $\mathcal{T}_{23}=2a_{2}m_{s}+4a_{3}m_{s}$, and the constants $a_{1}m_{s}$ and $a_{2}m_{s}$ are equal to $-67~{\rm MeV}$ and $134~{\rm MeV}$, respectively. In the aforementioned formula the strangeness content of the proton has a significant impact as it is directly related with the coefficient $a_{3}m_{s}$. Specifically, the allowed values for $a_{3}m_{s}$ are defined within the region $[-134,-310]~{\rm MeV}$, corresponding to strangeness of the order of $[0\%,20\%]$, respectively. Finally, the $\eta(x)$ function is a Heaviside function related to the appearance of muons. The contribution of leptons (electrons and muons) manifest as the relativistic expressions:
\begin{align}
    \tilde{\mathcal{E}}_{e}&=-\frac{\mu^{4}}{12\pi^{2} (\hbar c)^{3}}, \\
    \tilde{\mathcal{E}}_{\mu}&=\frac{m_{\mu}^{4}c^{8}}{8\pi^{2} (\hbar c)^{3}}\left[\left(2t^{2}+1\right)t\sqrt{t^{2}+1}-\sinh^{-1}t\right]\nonumber\\
    &- \mu \frac{p_{F_{\mu}}^{3}}{3\pi^{2} (\hbar c)^{3}},
\end{align}
where $p_{F_{\mu}}=\sqrt{\mu^{2}-m_{\mu}^{2}c^{4}}$ is the muon Fermi momentum and $t=p_{F_{\mu}}/(m_{\mu}c^{2})$.

The properties of NS matter, as they appear in the energy density, are determined by calculating the ground state of the effective energy density provided in Eq.~\eqref{eq:eff_energy_density_mdi}. This calculation involves extremizing the effective energy density for a fixed baryon state with respect to the proton fraction, chemical potential, and amplitude:
\begin{align}
    \frac{\partial \tilde{{\cal E}}}{\partial x}&: \mu - 4(1-2x)S(u)\sec^2\frac{\theta}{2}+2a_1m_s\tan^2\frac{\theta}{2} = 0, \nonumber \\
    \frac{\partial \tilde{{\cal E}}}{\partial \mu}&: \frac{f^2\mu}{(\hbar c)^{3}}\sin^2\theta+u\rho_0(1+x)\sin^2\frac{\theta}{2}-xu\rho_0+ \frac{\mu^3}{3\pi^2 (\hbar c)^{3}} \nonumber \\ &+\eta(|\mu|-m_{\mu} c^{2})\frac{(\mu^2-m_{\mu}^2 c^{4})^{3/2}}{3\pi^2 (\hbar c)^{3}}=0, \nonumber \\
    \frac{\partial \tilde{{\cal E}}}{\partial \theta}&: \cos\theta - \frac{1}{\mu^2}\left( m_K^2 c^{4}-\frac{(\hbar c)^{3} \mu}{2f^2}u\rho_0(1+x) \right. \nonumber \\ &+ \left. \frac{(\hbar c)^{3} u\rho_0}{2f^2}(2a_1x+2a_2+4a_3)m_s\right) = 0.
    \label{eq:eff_energy_system}
\end{align}
Eq.~\eqref{eq:eff_energy_system} provides the threshold density, the proton fraction and the chemical potential for the NS matter.

\subsubsection{Equation of State}
Utilizing the quantities derived from Eq.~\eqref{eq:eff_energy_system}, the energy density for the construction of the EoS is given by the relation~\cite{Thorsson-1994}:
\begin{align}
    {\cal E}(u,x,\mu,\theta) &= {\cal E}_{\rm MDI}(u,x=1/2) + u\rho_{0} (1-2x)^{2} S(u) \nonumber \\ 
    &+ f^{2}\frac{\mu^{2}}{2 (\hbar c)^{3}}\sin^{2}\theta+f^{2}\frac{2m_K^{2} c^{4}}{(\hbar c)^{3}}\sin^2\frac{\theta}{2} \nonumber \\
    &+ (2a_{1}m_{s}x+\mathcal{T}_{23})u \rho_{0}\sin^{2}\frac{\theta}{2} \nonumber\\
    &+ {\cal E}_{e}+\eta(|\mu|-m_{\mu} c^{2}){\cal E}_{\mu},
    \label{eq:energy_density}
\end{align}
where:
\begin{align}
    \mathcal{E}_{e}&=\frac{\mu^{4}}{4\pi^{2} (\hbar c)^{3}}, \\
    \mathcal{E}_{\mu}&=\frac{m_{\mu}^{4} c^{8}}{8\pi^{2} (\hbar c)^{3}}\left[\left(2t^{2}+1\right)t\sqrt{t^{2}+1} - \sinh^{-1}{t}\right],
\end{align}
and the pressure is given by the relation~\cite{Moustakidis-2007}:
\begin{align}
    p(u,x,\mu,\theta) &= p_{b}(u,x) + p_{K}(\mu,\theta) \nonumber \\
    &+ p_{e} + \eta(|\mu|-m_{\mu}c^{2})p_{\mu},
\end{align}
where:
\begin{align}
    p_{b}(u,x) &= u^{2} \frac{\partial}{\partial u} \left(\frac{\mathcal{E}_{b}(u,x)}{u}\right), \nonumber \\
    p_{K}(\mu,\theta) &= -f^2\frac{\mu^2}{2 (\hbar c)^{3}}\sin^2\theta-f^{2}\frac{2m_K^2 c^{4}}{(\hbar c)^{3}}\sin^2\frac{\theta}{2}, \nonumber \\
    p_{l=e,\mu} &= \frac{m_{l}^{4}c^{5}}{24\pi^{2}\hbar^{3}}\left[\left(2z^{3}-3z\right)\left(1+z^{2}\right)^{1/2} + 3\sinh^{-1}{z}\right],
\end{align}
with $z = \left(\hbar c\right)\left(3\pi^{2} \rho x_{l}\right)^{1/3} (m_{l}c^{2})^{-1}$, and $x_{l}$ being the lepton fraction.

\subsection{Neutron star structure}
The mechanical equilibrium of the NS matter is determined by the system of two differential equations, the well known Tolman–Oppenheimer–Volkoff (TOV) equations, solved in a self-consistent way with the EoS of the fluid interior~\cite{Shapiro-1983,Glendenning-2000}. The formulae for the TOV equations are given as:
\begin{align}
	\frac{dP(r)}{dr}&=-\frac{G\rho(r) M(r)}{r^2}\left(1+\frac{P(r)}{\rho(r) c^2}\right)\left(1+\frac{4\pi P(r) r^3}{M(r)c^2}\right) \\\nonumber & \left(1-\frac{2GM(r)}{c^2r}\right)^{-1}, \\
    \frac{dM(r)}{dr}&=4\pi r^2\rho(r).
\end{align}
By solving this system of differential equations, we determine the fundamental properties of the NS, such as its mass $M$ and radius $R$, which are influenced by the applied EoS.

In addition, an important and well measured quantity by the gravitational wave detectors, which can be treated as a tool to impose constraints on the EoS, is the dimensionless tidal deformability $\Lambda$, defined as
\begin{equation}
    \Lambda=\frac{2}{3}k_2 \left(\frac{c^2R}{GM}\right)^5,
\end{equation}
where $k_2$  is the tidal Love number \cite{Flanagan-08,Hinderer-08}. As $\Lambda$ is sensitive to the NS radius, it can provide information for the low density part of the EoS.

For the construction of the crust EoS we employ the well known model of Baym, Pethick, and Sutherland~\cite{Baym-71}.

\section{Results and Implications}
\label{sec:Results_and_Implications}
\subsection{EoSs: Pressure - Density relations}
The EoS for the RMF+FOKC model is based on the EoS E1 from Ref.~\cite{Kanakis-Petousis-2024} since it fulfilled the maximum mass constraints~\cite{Arzoumanian-2018,Antoniadis-2013,Cromartie-2020,Romani-2022} and the ones derived from GW170817 event~\cite{Abbott-2019}. Since in Ref.~\cite{Kanakis-Petousis-2024} an admixture of hypothetical 17 MeV boson was introduced and the resulting effective mass of vector boson was used, the parameters of the EoS were re-scaled back to the $\omega$ meson  mass, while the results remained the same. This is possible since the RMF theory depends on ratios of meson mass and corresponding coupling and the same EoS can be expressed by various sets of parameters. The parameters from Ref.~\cite{Kanakis-Petousis-2024} and re-scaled ones are shown in Table~\ref{tab:tabeos}.

\begin{table}[b]
\caption{Parameter sets for RMF EoS. EoS E1 corresponds to the EoS extracted in Ref.~\cite{Kanakis-Petousis-2024} while EoS E1-K to the values re-scaled to the $\omega$-meson mass. Coupling for $\rho$-meson is $g_{\rho N}=4.69$ in both cases.}
\begin{center}
\vspace{0.2cm}
    \begin{tabular}{lrrr}
    \hline
    Parameters & EoS E1 & EoS E1-K & Units \\
    \hline
        m$_{v}$ & 626.00 & 783.00 & MeV \\
        m$_{s}$ & 406.60 & 508.57 & MeV \\
        g$_{v}$ & 7.61 & 9.52 & \\
        g$_{s}$ & 6.78 & 8.48 \\
        $\kappa$ & 19.00 & 37.18 & MeV \\
        $\lambda$ & -60.00 & -146.86 &  \\
    \hline
    \end{tabular}
\end{center}
\label{tab:tabeos}
\end{table}

\begin{figure}
\centering
\includegraphics[trim=.45cm .1cm 1.5cm 1cm, width=\columnwidth, clip, height = 7.75cm]{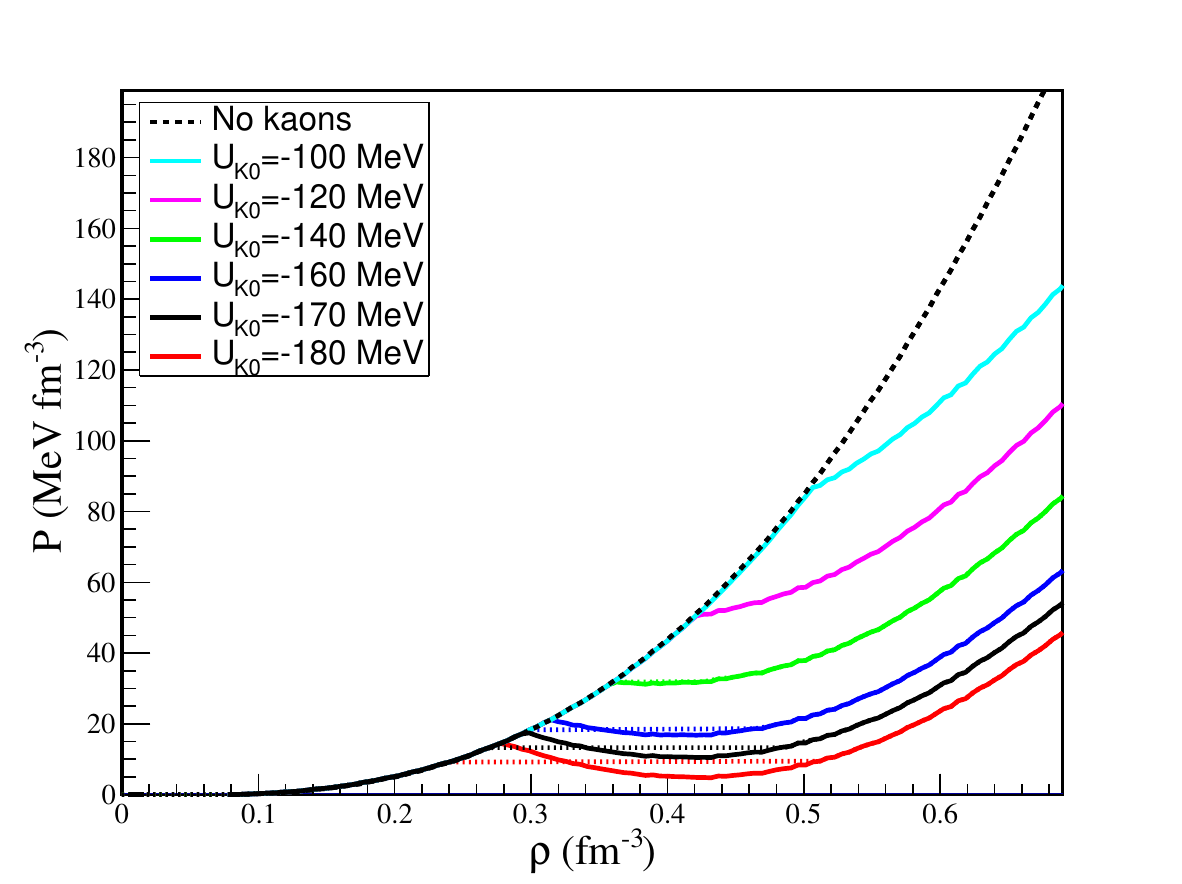}
\caption{Pressure as a function of density for the E1 EoS and the corresponding E1-K EoSs with several kaon potentials U$_{K0}$=-180, -170, -160, -140, -120, -100 MeV. The normal matter is presented with the dashed line, while the exotic matter with the solid lines. Horizontal dotted lines denote the corresponding Maxwell construction.}
\label{fig:fgkeos}
\end{figure}

Fig.\ref{fig:fgkeos} displays the pressure as a function of the density for the E1-K EoS with kaon potentials $U_{K0}=-180,-170,-160,-140,-120,-100~{\rm MeV}$. It is apparent that with stronger kaon potential (lower values of $U$) a shallow dip in the pressure appears after the threshold value for the baryonic density. Such behavior influences the properties of a compact object. 

Considering the $\chi$-MDI model, the parametrization for the MDI+APR1 EoS~\cite{Koliogiannis-2021}, which meets the maximum mass constraints~\cite{Arzoumanian-2018,Antoniadis-2013,Cromartie-2020,Romani-2022} and the ones derived from GW170817 event~\cite{Abbott-2019}, was employed regarding the hadronic matter. In analogy to the RMF+FOKC model, the term that is related to the appearance of kaons is the parameter $\alpha_{3}m_{s}$, explicitly connected to the kaon-nucleon sigma term~\cite{Thorsson-1994}. Although the boundaries for the $\alpha_{3}m_{s}$ are $[-134,-310]~{\rm MeV}$, the values that we utilize in the present study are from $-238~{\rm MeV}$ up to $-280~{\rm MeV}$. The limits were selected as boundary conditions between the appearance of a kaon condensation and the HESS J1731-347 1$\sigma$ region.

\begin{figure}
\includegraphics[width=\columnwidth]{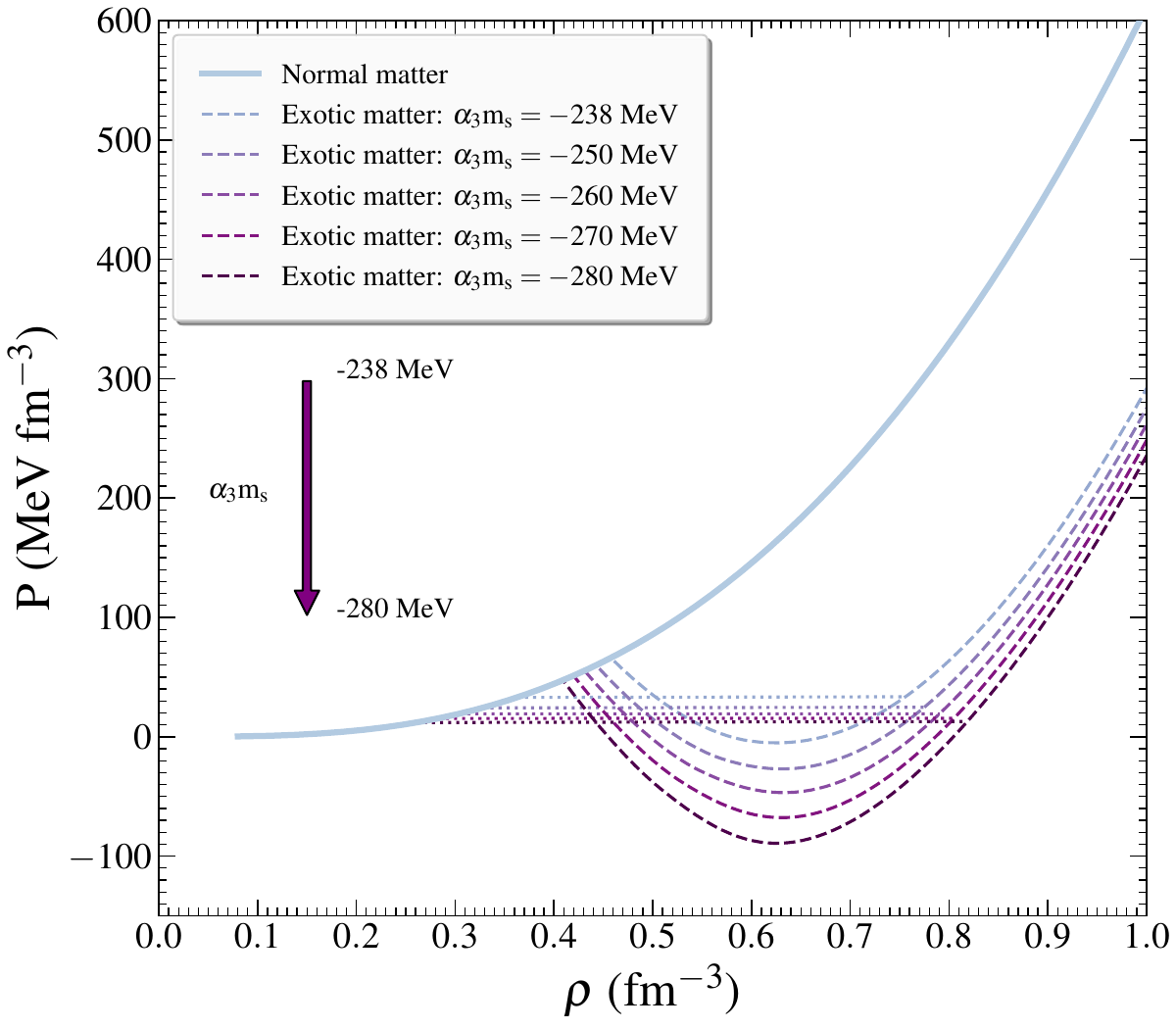}
\caption{Pressure as a function of density for the MDI+APR1 EoS and the corresponding EoSs with $\alpha_{3}m_{s}$ in the range $[-238,-280]~{\rm MeV}$. The normal matter is presented with the solid line, while the exotic matter with the dashed lines. The arrow indicates the increase in the strangeness content of the proton. Horizontal dotted lines denote the corresponding Maxwell construction.}
\label{fig:energy_pressure}
\end{figure}

Fig.~\ref{fig:energy_pressure} indicates the pressure as a function of density for the MDI+APR1 EoS and the corresponding EoSs with $a_{3}m_{s}$ in the range $[-238,-280]~{\rm MeV}$. As the strangeness content of the proton increases, or the $a_{3}m_{s}$ decreases, the threshold density for the appearance of a kaon condensate decreases. Additionally, the aforementioned increase leads to decrease in pressure following the threshold density. This reduction is becoming increasingly significant as the content of the strangeness of the proton is increasing. 

The treatment for the decrease in pressure in both models, since the condensation is strong in NSs, is featuring the Maxwell construction (MC) for the existence of a single value of chemical potential and pressure. MC is presented in Figs.~\ref{fig:fgkeos} and~\ref{fig:energy_pressure} with the horizontal lines.

\subsection{Proton fraction and speed of sound}
\begin{figure}
\includegraphics[trim=.45cm .1cm 1.5cm 1cm, width=\columnwidth, clip, height = 7.75cm]{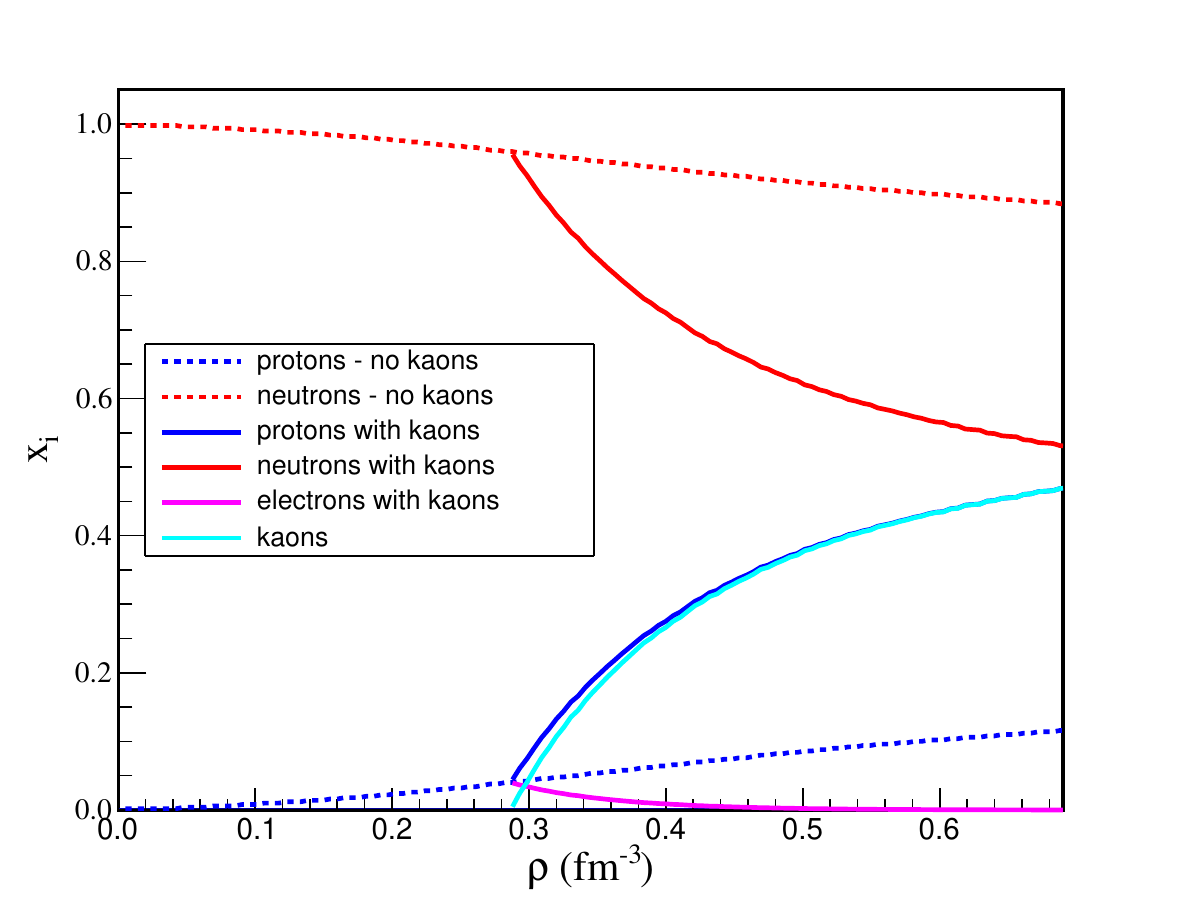}
\caption{Particle fractions as a function of the density for the E1 and E1-K EoSs with kaon potential U$_{K0}$=-170 MeV.}
\label{fig:pf_rmf}
\end{figure}

\begin{figure}[h]
\includegraphics[width=\columnwidth]{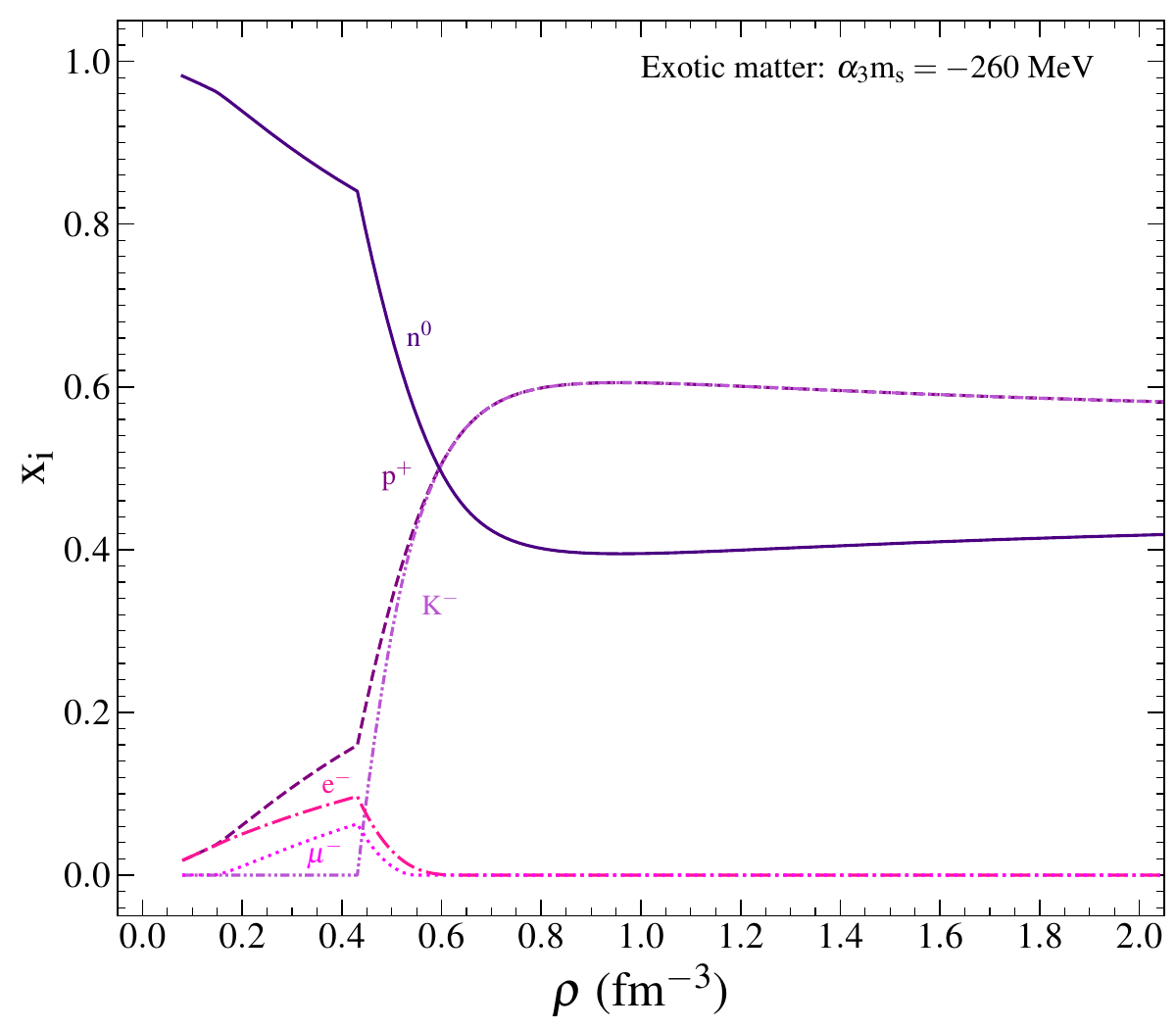}
\caption{Particle fractions as a function of the density for the MDI+APR1 EoS with $\alpha_{3}m_{s}=-260~{\rm MeV}$. Neutrons are depicted using a solid line, protons by a dashed line, electrons by a dash-dotted line, muons by a dotted line, and kaons by a dash-dot-dotted line.}
\label{fig:pf_mdi}
\end{figure}

In the structure of NSs major role possesses the proton fraction, as well as the fractions of the rest particles that have been considered in the $\beta$-equilibrium state. Figs.~\ref{fig:pf_rmf} and~\ref{fig:pf_mdi} present the fraction of the particles as a function of the density throughout the star for the onset of a kaon condensation for the E1-K and MDI+APR1 EoSs, respectively. Both figures indicate only one scenario for each case (similar results are produced for the rest scenarios). In all cases, the net negative charge in the kaon fields causes the decrease in the leptons concentration. Furthermore, for densities near and higher than the threshold density, the fraction of kaons increases rapidly and eventually replace the negative leptons in NS matter.

\begin{figure}[h!]
\includegraphics[width=\columnwidth]{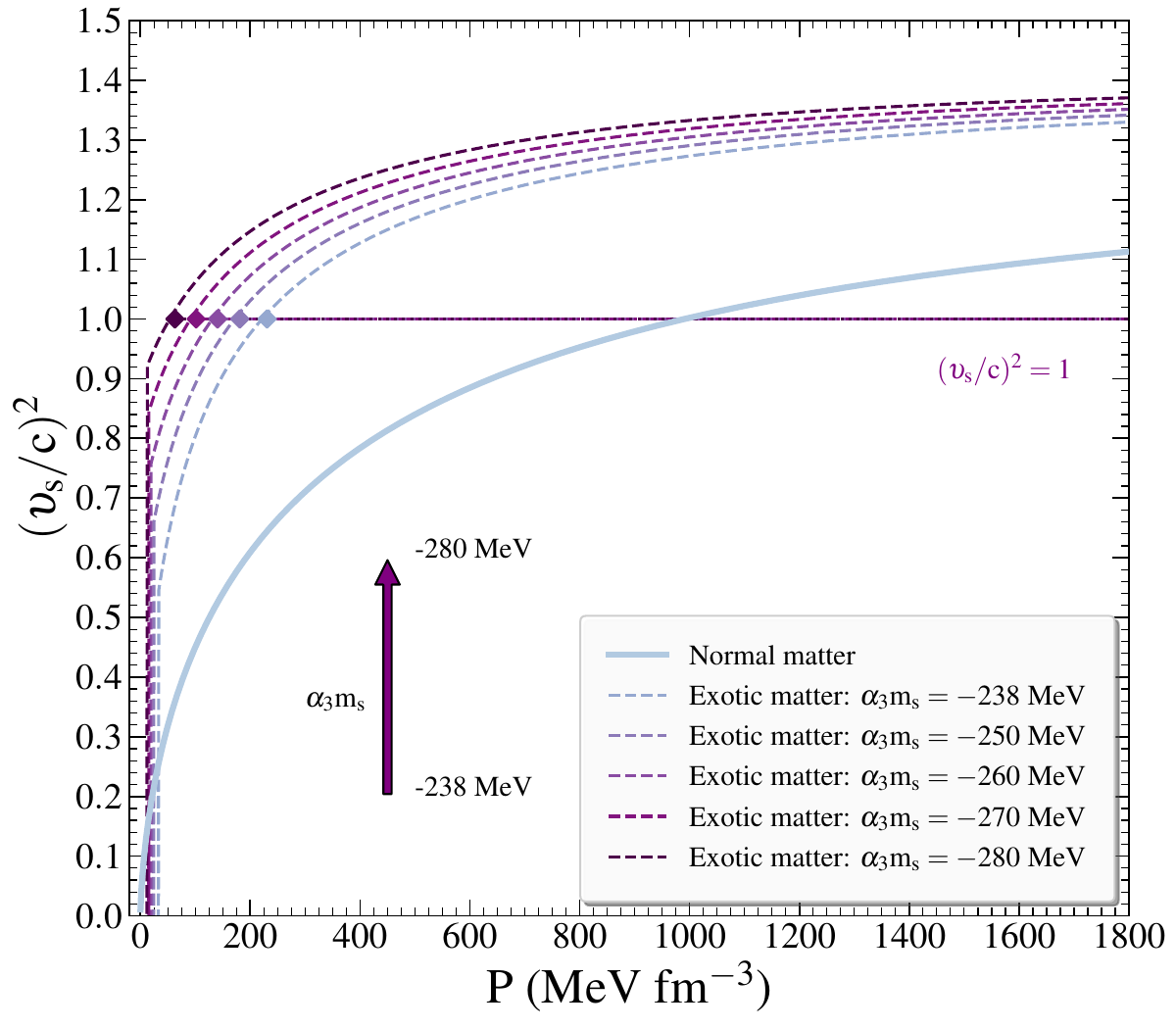}
\caption{Square speed of sound in units of speed of light as a function of the pressure for the MDI+APR1 EoS and the corresponding EoSs with $\alpha_{3}m_{s}$ in the range $[-238,-280]~{\rm MeV}$. The EoS featuring normal matter is depicted by the solid line, whereas the EoSs with exotic matter are represented by the dashed lines. Diamonds indicate the pressure at which $(\upsilon_{s}/c)=1$. The dotted lines present the EoSs with MC considering the maximum allowed value in the speed of sound. The arrow indicates the increase in the strangeness content of the proton.}
\label{fig:sos_mdi}
\end{figure}

Moreover, another microscopic quantity that is related to the structure of the star, is the speed of sound, which must not exceed the causality limit at least up to densities that correspond to maximum mass configuration. This statement is always fulfilled in the case of the RMF model as it is a relativistic approach. In the MDI model, which is a non-relativistic model, the validity of the speed of sound must be checked. For that reason, in Fig.~\ref{fig:sos_mdi} we depict the square speed of sound in units of speed of light as a function of the pressure for the MDI+APR1 EoS and the corresponding EoSs with $a_{3}m_{s}$ in the range $[-238,-280]~{\rm MeV}$. In all cases, the onset of the kaon condensate violates the causality limit. It needs to be noted that in the case of hadronic MDI+APR1 EoS, while the speed of sound violates the causality limit, this effect take place at higher densities than the maximum mass configuration~\cite{Koliogiannis-2021}. To treat the acausal behavior of the EoS with the onset of kaons, we employed a MC at the density where $(\upsilon_{s}/c)=1$. Beyond that density we introduced the maximally stiff EoS with $(\upsilon_{s}/c)=1$.

\subsection{Mass-Radius plane}
\begin{figure*}[t!]
\begin{overpic}[trim=.45cm 0.3cm 0.8cm 0.17cm, width=\columnwidth, clip, height = 7.4cm]{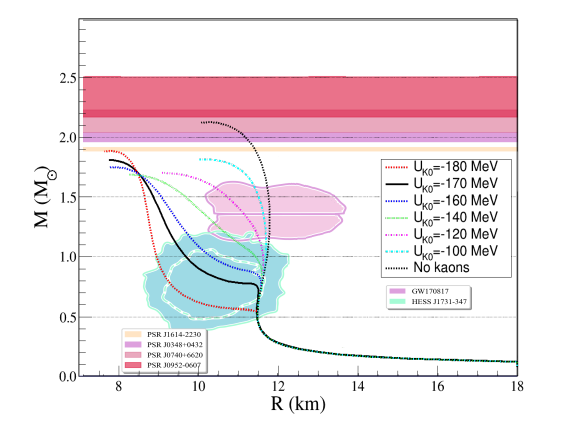}
 \put (15,75) {\scriptsize (a)}
\end{overpic}
\begin{overpic}[trim=.0cm .0cm 1.6cm 1.2cm, width=\columnwidth, clip, height = 7.4cm]{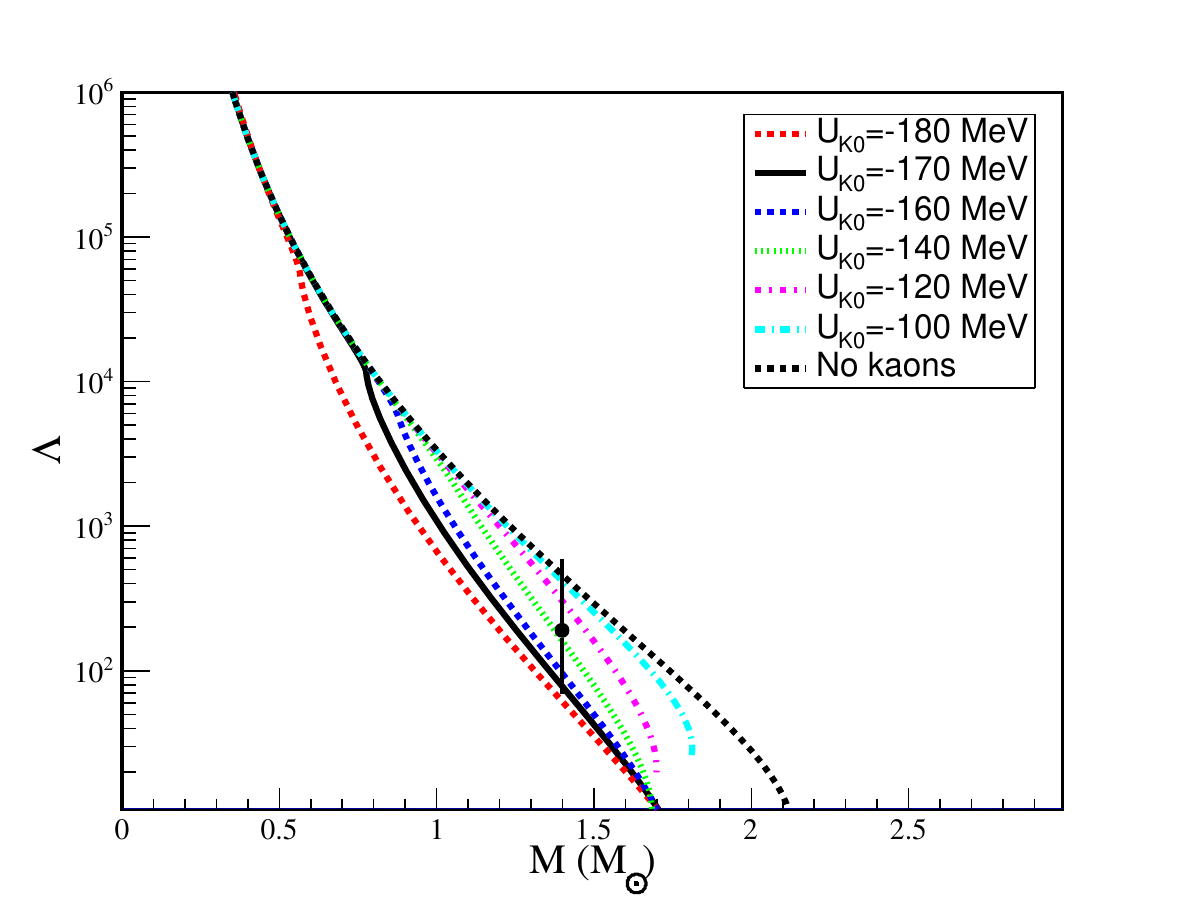}
 \put (15,75) {\scriptsize (b)}
\end{overpic}
\caption{(a) Mass radius diagram for RMF EoS and several kaon potentials (U$_{K0}$=-180, -170, -160, -140, -120, -100 MeV) and without kaons. The shaded regions from bottom to top represent the HESS J1731-347 remnant~\cite{Doroshenko-2022}, the GW170817 event~\cite{Abbott-2019}, and the PSR J1614-2230~\cite{Arzoumanian-2018}, PSR J0348+0432~\cite{Antoniadis-2013}, PSR J0740+6620~\cite{Cromartie-2020}, and PSR J0952-0607~\cite{Romani-2022} pulsar observations with possible maximum NS mass. (b) The corresponding dimensionless tidal deformability as a function of the gravitational mass. The point at $\rm 1.4~M_{\odot}$ indicates the constraints from the GW170817 event~\cite{Abbott-2019}.}
\label{fig:fgmr}
\end{figure*}

Fig.~\ref{fig:fgmr}\textcolor{blue}{(a)} presents the gravitational mass as a function of the radius for the E1-K EoS with kaon potentials $U_{K0}=[-180,-170,-160,-140,-120,-100]~{\rm MeV}$. The values of couplings $g_{\omega K}$ and $g_{\rho k}$ were taken as in Ref.~\cite{Glendenning-1999} ($g_{\omega K}=g_{\omega N}/3$, $g_{\rho K} = g_{\rho N}$). In case of the three strongest kaon potentials considered, $U_{K0}=[-180,-170,-160]~{\rm MeV}$, the curves crosses the region of the reported HESS J1731-347~\cite{Doroshenko-2022} compact object. Specifically, the case of $U_{K0}=-170~{\rm MeV}$, as an intermediate value, lies closer to the centre of the object. Thus, HESS J1731-347 can be explained as a NS with kaon condensate, which forms an alternative branch of the mass-radius diagram.

\begin{figure*}[t!]
\includegraphics[width=\columnwidth]{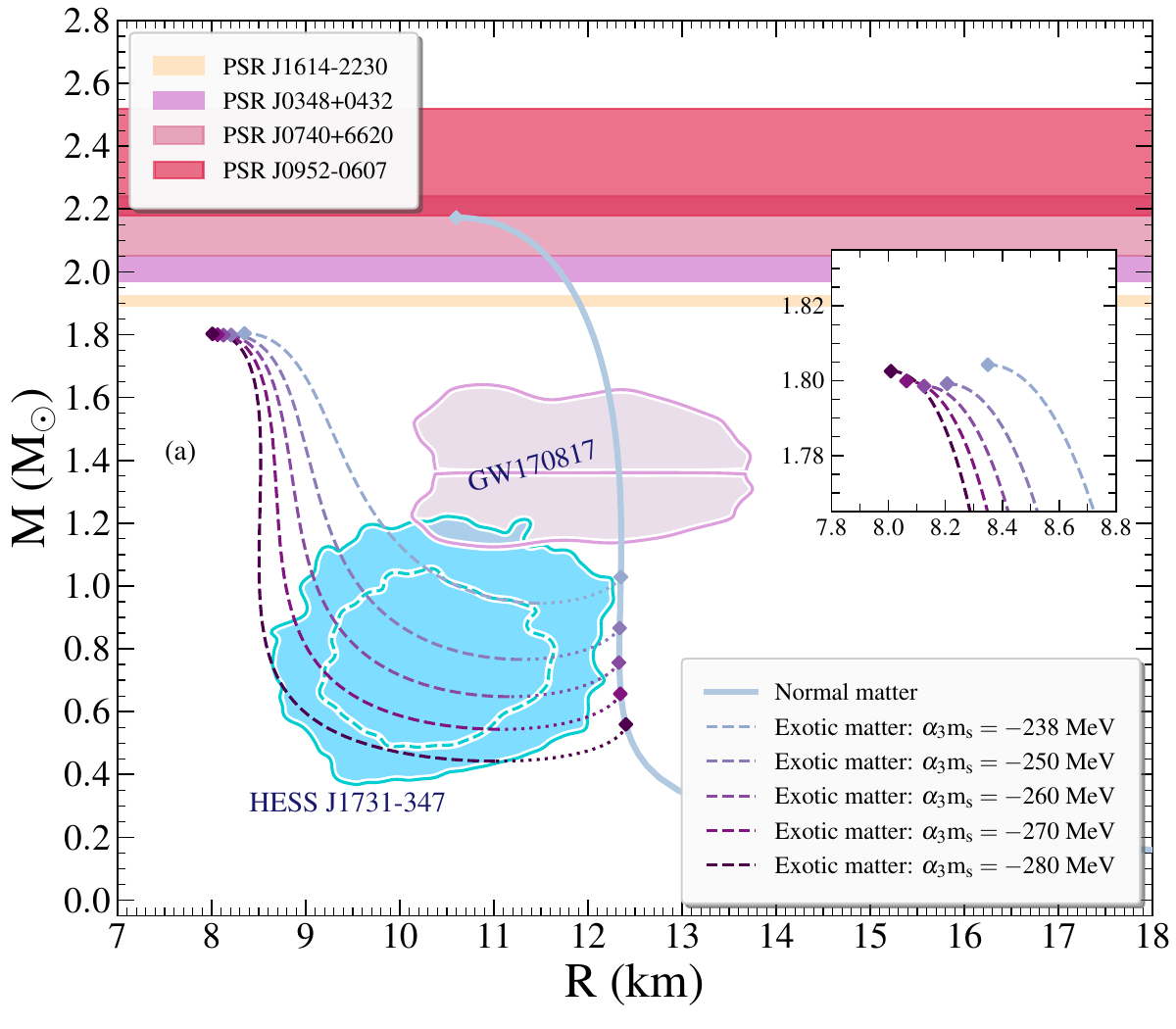}
~
\includegraphics[width=\columnwidth]{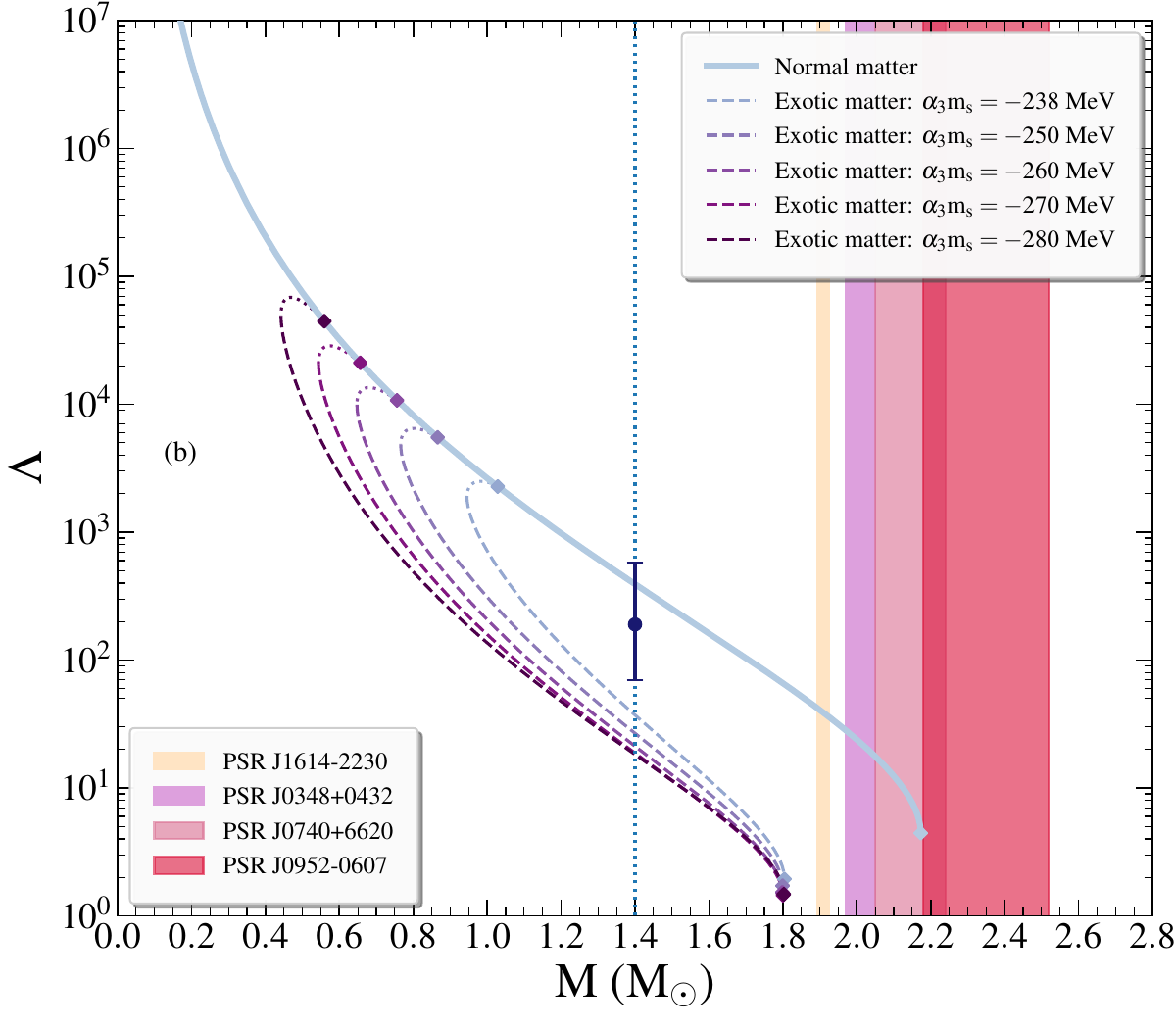}
\caption{(a) Gravitational mass as a function of the radius for the MDI+APR1 EoS and the corresponding EoSs with $\alpha_{3}m_{s}$ in the range $[-238,-280]~{\rm MeV}$. The EoS featuring normal matter is depicted by the solid line, whereas the EoSs with exotic matter are represented by the dashed lines. Diamonds indicate the maximum mass configurations. The shaded regions from bottom to top represent the HESS J1731-347 remnant~\cite{Doroshenko-2022}, the GW170817 event~\cite{Abbott-2019}, and the PSR J1614-2230~\cite{Arzoumanian-2018}, PSR J0348+0432~\cite{Antoniadis-2013}, PSR J0740+6620~\cite{Cromartie-2020}, and PSR J0952-0607~\cite{Romani-2022} pulsar observations with possible maximum NS mass. (b) The corresponding dimensionless tidal deformability as a function of the gravitational mass. The point at $\rm 1.4~M_{\odot}$ indicates the constraints from the GW170817 event~\cite{Abbott-2019}.}
\label{fig:mass_radius_mdi}
\end{figure*}

In the scenario of MDI model, the MDI+APR1 EoS and the corresponding EoSs with $a_{3}m_{s}$ in the range $[-238,-280]~{\rm MeV}$ as the dependence of gravitational mass on the radius are displayed in Fig.~\ref{fig:mass_radius_mdi}\textcolor{blue}{(a)}. The onset of a kaon condensation in NSs manifest a strong softening on the hadronic EoS by reducing its maximum mass at around 16\%. It is evident that these stars explain remarkably accurate the ultralight compact object within the supernova remnant HESS J1731-347. Thus, the observed ultralight compact object enhances the hypothesis that is rather a star with exotic core than the lightest NS~\cite{Doroshenko-2022}. In addition, this softening leads to a third gravitationally stable branch on the mass-radius plane giving rising to twin stars. The new stable branch results in stars with the same mass as normal compact stars but quite different radii. Therefore, a kaon condensation provides an additional way for the appearance of twin stars. 

At this point, it is worth noting the recent reports of several light compact objects, including HESS J1731-347, XTE J1814-338~\cite{10.1093/mnras/stae2398} ($M = 1.21^{+0.05}_{–0.05}~{\rm M_{\odot}}$ and $R=7^{+0.4}_{-0.4}~{\rm km}$), and PSR J1231-1411~\cite{Salmi_2024} ($M = 1.04^{+0.05}_{–0.03}~{\rm M_{\odot}}$ and $R=12.6^{+0.3}_{-0.3}~{\rm km}$). These observations highlight the challenge of simultaneously describing such objects while satisfying the constraints on the maximum mass of neutron stars. This diversity in light compact objects evidently necessitates a corresponding diversity in the underlying physical scenarios, each characterized by a distinct structural composition. In particular, strangeness may serve as an additional degree of freedom, the extent of which could be determined by the formation scenario of each object.

As in nuclear reactions, the production of strangeness requires higher densities and energies (or temperatures). Consequently, compact objects containing strangeness are likely formed during the collapse of more massive stellar progenitors, where the extreme conditions facilitate the copious production of strangeness. Within this framework, our model can be interpreted as consisting of two branches: a nucleonic branch, capable of reproducing the maximum mass of neutron stars, and a kaonic-condensate branch, which can reproduce the properties of light compact objects, such as HESS J1731-347. Furthermore, by introducing additional condensate, this framework may also describe objects like XTE J1814-338. This hypothesis will be investigated in a future work.

For completeness, Figs.~\ref{fig:fgmr}\textcolor{blue}{(b)} and~\ref{fig:mass_radius_mdi}\textcolor{blue}{(b)} display the dimensionless tidal deformability as a function of the gravitational mass. E1-K EoS containing the kaon condensation originated from the RMF model fulfills, in the majority of the potentials, the constraint extracted through the gravitational wave event GW170817~\cite{Abbott-2019}, while the MDI+APR1 EoSs with the onset of kaons based on the MDI model, lying outside the defined boundaries. 

\section{Remarks}
\label{sec:Remarks}
The observation of the ultralight compact object within the supernova remnant HESS J1731-347 enhanced the speculation of the existence of exotic core in NSs. In this direction, we present a first attempt aiming to describe the ultralight compact object within the framework of a kaon condensate in nuclear matter. The main conjecture was that the appearance of a kaon condensation in dense nuclear matter would be able to soften the hadronic EoS appropriately to describe the aforementioned compact object. Two theoretical frameworks for the hadronic matter were employed, where the onset of kaons is introduced in a different way. 

The aforementioned procedure is utilized to cover different nuclear models and their implications on the kaon condensate. The onset of kaon condensate on both theoretical nuclear models describes accurately the ultralight compact object within the supernova remnant HESS J1731-347. Additionally, a notable feature emerges in the form of a knot around $1.8~{\rm M_{\odot}}$ and $8~{\rm km}$. This convergence suggests an underlying universality among the models, supporting the hypothesis that detections of stars at this specific point could correspond to stars with the onset of kaon condensation. At this point, the mass-radius relations for different values of kaon potentials or the parameter $\alpha_{3}m_{s}$ intersect, indicating a significant effect on the structure of the compact object. This may signal a subtle balance where earlier onset of kaon condensation, associated with stiffer kaon potentials, is compensated by softer nucleonic EoS that supports less baryonic matter. Should a compact object detected at this point in the mass-radius plane in the future, it could be attributed to a NS with a kaon condensate, independent of the specific model of kaon condensation employed.

Furthermore, the appearance of exotic matter in the form of a kaon condensate into a zero momentum state, strengthens the hypothesis of twin stars. Concluding, an exotic core in NS as a condensate of negatively charged kaons, reduces sufficiently the energy density of dense nuclear matter and leads to accurate description of ultralight compact objects, such as the HESS J1731-347.

As a general consideration, the hypothesis that strange kaonic matter is present in the ultralight compact object within the supernova remnant HESS J1731-347 may depend on the astrophysical processes involved in its formation. Possible scenarios for such formation have been suggested, for example, in Ref.~\cite{KaonReview}.

Future observations of similar to HESS J1731-347 object will significantly improve our knowledge for the currently unknown behavior of superdense nuclear matter by imposing valuable constraints on its properties.

\section*{Acknowledgments}
This work is supported by the Croatian Science Foundation under the project number HRZZ-MOBDOL-12-2023-6026 and under the project Relativistic Nuclear Many-Body Theory in the Multimessenger Observation Era (IP-2022-10-7773), and by the Czech Science Foundation (GACR Contract No. 21-24281S). 
\bibliographystyle{elsarticle-num}
\bibliography{bibliography}

\begin{thebibliography}{10}
\expandafter\ifx\csname url\endcsname\relax
  \def\url#1{\texttt{#1}}\fi
\expandafter\ifx\csname urlprefix\endcsname\relax\def\urlprefix{URL }\fi
\expandafter\ifx\csname href\endcsname\relax
  \def\href#1#2{#2} \def\path#1{#1}\fi

\bibitem{Griffin_Snoke_Stringari_1995}
A.~Griffin, D.~W. Snoke, S.~Stringari (Eds.), Bose-{E}instein {C}ondensation,
  Cambridge University Press, 1995.

\bibitem{Brown-1992}
G.~Brown, K.~Kubodera, M.~Rho, V.~Thorsson,
  \href{https://www.sciencedirect.com/science/article/pii/037026939291386N}{A
  novel mechanism for kaon condensation in neutron star matter}, Phys. Lett. B
  291~(4) (1992) 355--362.
\newblock \href {https://doi.org/https://doi.org/10.1016/0370-2693(92)91386-N}
  {\path{doi:https://doi.org/10.1016/0370-2693(92)91386-N}}.
\newline\urlprefix\url{https://www.sciencedirect.com/science/article/pii/037026939291386N}

\bibitem{Thorsson-1994}
V.~Thorsson, M.~Prakash, J.~M. Lattimer,
  \href{https://www.sciencedirect.com/science/article/pii/0375947494904073}{Composition,
  structure and evolution of neutron stars with kaon condensates}, Nuc. Phys. A
  572~(3) (1994) 693--731.
\newblock \href {https://doi.org/https://doi.org/10.1016/0375-9474(94)90407-3}
  {\path{doi:https://doi.org/10.1016/0375-9474(94)90407-3}}.
\newline\urlprefix\url{https://www.sciencedirect.com/science/article/pii/0375947494904073}

\bibitem{Glendenning-1998}
N.~K. Glendenning, J.~Schaffner-Bielich,
  \href{https://link.aps.org/doi/10.1103/PhysRevLett.81.4564}{Kaon
  {C}ondensation and {D}ynamical {N}ucleons in {N}eutron {S}tars}, Phys. Rev.
  Lett. 81 (1998) 4564--4567.
\newblock \href {https://doi.org/10.1103/PhysRevLett.81.4564}
  {\path{doi:10.1103/PhysRevLett.81.4564}}.
\newline\urlprefix\url{https://link.aps.org/doi/10.1103/PhysRevLett.81.4564}

\bibitem{Glendenning-1999}
N.~Glendenning, J.~Schaffner-Bielich, First order kaon condensate., Phys. Rev.
  C 60 (1999) 025803.
\newblock \href {https://doi.org/https://doi.org/10.1103/PhysRevC.60.025803}
  {\path{doi:https://doi.org/10.1103/PhysRevC.60.025803}}.

\bibitem{Lim-2014}
Y.~Lim, K.~Kwak, C.~H. Hyun, C.-H. Lee,
  \href{https://link.aps.org/doi/10.1103/PhysRevC.89.055804}{Kaon condensation
  in neutron stars with {S}kyrme-{H}artree-{F}ock models}, Phys. Rev. C 89
  (2014) 055804.
\newblock \href {https://doi.org/10.1103/PhysRevC.89.055804}
  {\path{doi:10.1103/PhysRevC.89.055804}}.
\newline\urlprefix\url{https://link.aps.org/doi/10.1103/PhysRevC.89.055804}

\bibitem{PhysRevD.102.123007}
V.~B. Thapa, M.~Sinha,
  \href{https://link.aps.org/doi/10.1103/PhysRevD.102.123007}{Dense matter
  equation of state of a massive neutron star with antikaon condensation},
  Phys. Rev. D 102 (2020) 123007.
\newblock \href {https://doi.org/10.1103/PhysRevD.102.123007}
  {\path{doi:10.1103/PhysRevD.102.123007}}.
\newline\urlprefix\url{https://link.aps.org/doi/10.1103/PhysRevD.102.123007}

\bibitem{Menezes-2005}
D.~Menezes, P.~Panda, C.~Providencia, Kaon condensation in the quark-meson
  coupling model and compact stars., Phys. Rev. C 72 (2005) 035802.
\newblock \href {https://doi.org/https://doi.org/10.1103/PhysRevC.72.035802}
  {\path{doi:https://doi.org/10.1103/PhysRevC.72.035802}}.

\bibitem{Kaplan-1986}
D.~Kaplan, A.~Nelson,
  \href{https://www.sciencedirect.com/science/article/pii/037026938690331X}{Strange
  goings on in dense nucleonic matter}, Phys. Lett. B 175~(1) (1986) 57--63.
\newblock \href {https://doi.org/https://doi.org/10.1016/0370-2693(86)90331-X}
  {\path{doi:https://doi.org/10.1016/0370-2693(86)90331-X}}.
\newline\urlprefix\url{https://www.sciencedirect.com/science/article/pii/037026938690331X}

\bibitem{Brown-1994}
G.~Brown, C.-H. Lee, M.~Rho, V.~Thorsson,
  \href{https://www.sciencedirect.com/science/article/pii/0375947494903352}{From
  kaon-nuclear interactions to kaon condensation}, Nuc. Phys. A 567~(4) (1994)
  937--956.
\newblock \href {https://doi.org/https://doi.org/10.1016/0375-9474(94)90335-2}
  {\path{doi:https://doi.org/10.1016/0375-9474(94)90335-2}}.
\newline\urlprefix\url{https://www.sciencedirect.com/science/article/pii/0375947494903352}

\bibitem{PhysRevD.103.063004}
V.~B. Thapa, M.~Sinha, J.~J. Li, A.~Sedrakian,
  \href{https://link.aps.org/doi/10.1103/PhysRevD.103.063004}{Massive
  {$\mathrm{\ensuremath{\Delta}}$}-resonance admixed hypernuclear stars with
  antikaon condensations}, Phys. Rev. D 103 (2021) 063004.
\newblock \href {https://doi.org/10.1103/PhysRevD.103.063004}
  {\path{doi:10.1103/PhysRevD.103.063004}}.
\newline\urlprefix\url{https://link.aps.org/doi/10.1103/PhysRevD.103.063004}

\bibitem{PhysRevC.105.015807}
F.~Ma, W.~Guo, C.~Wu,
  \href{https://link.aps.org/doi/10.1103/PhysRevC.105.015807}{Kaon meson
  condensate in neutron star matter including hyperons}, Phys. Rev. C 105
  (2022) 015807.
\newblock \href {https://doi.org/10.1103/PhysRevC.105.015807}
  {\path{doi:10.1103/PhysRevC.105.015807}}.
\newline\urlprefix\url{https://link.aps.org/doi/10.1103/PhysRevC.105.015807}

\bibitem{PhysRevC.107.035807}
D.~Kundu, V.~B. Thapa, M.~Sinha,
  \href{https://link.aps.org/doi/10.1103/PhysRevC.107.035807}{(anti)kaon
  condensation in strongly magnetized dense matter}, Phys. Rev. C 107 (2023)
  035807.
\newblock \href {https://doi.org/10.1103/PhysRevC.107.035807}
  {\path{doi:10.1103/PhysRevC.107.035807}}.
\newline\urlprefix\url{https://link.aps.org/doi/10.1103/PhysRevC.107.035807}

\bibitem{SEDRAKIAN2023104041}
A.~Sedrakian, J.~J. Li, F.~Weber,
  \href{https://www.sciencedirect.com/science/article/pii/S0146641023000224}{Heavy
  baryons in compact stars}, Prog. Part. Nucl. Phys. 131 (2023) 104041.
\newblock \href {https://doi.org/https://doi.org/10.1016/j.ppnp.2023.104041}
  {\path{doi:https://doi.org/10.1016/j.ppnp.2023.104041}}.
\newline\urlprefix\url{https://www.sciencedirect.com/science/article/pii/S0146641023000224}

\bibitem{PhysRevD.107.074007}
C.~Adam, A.~G. Mart\'{\i}n-Caro, M.~Huidobro, A.~Wereszczynski, R.~V\'azquez,
  \href{https://link.aps.org/doi/10.1103/PhysRevD.107.074007}{Kaon condensation
  in skyrmion matter and compact stars}, Phys. Rev. D 107 (2023) 074007.
\newblock \href {https://doi.org/10.1103/PhysRevD.107.074007}
  {\path{doi:10.1103/PhysRevD.107.074007}}.
\newline\urlprefix\url{https://link.aps.org/doi/10.1103/PhysRevD.107.074007}

\bibitem{Sharifi_2021}
Z.~Sharifi, M.~Bigdeli, D.~Alvarez-Castillo, E.~Nasiri,
  \href{https://dx.doi.org/10.1088/1402-4896/ac30a5}{Binary neutron star
  mergers within kaon condensation:{GW}170817}, Phys. Scr. 96~(12) (2021)
  125311.
\newblock \href {https://doi.org/10.1088/1402-4896/ac30a5}
  {\path{doi:10.1088/1402-4896/ac30a5}}.
\newline\urlprefix\url{https://dx.doi.org/10.1088/1402-4896/ac30a5}

\bibitem{PhysRevC.108.045803}
S.~Kubis, W.~W\'ojcik, D.~A. Castillo, N.~Zabari,
  \href{https://link.aps.org/doi/10.1103/PhysRevC.108.045803}{Relativistic
  mean-field model for the ultracompact low-mass neutron star {HESS}
  {J}1731-347}, Phys. Rev. C 108 (2023) 045803.
\newblock \href {https://doi.org/10.1103/PhysRevC.108.045803}
  {\path{doi:10.1103/PhysRevC.108.045803}}.
\newline\urlprefix\url{https://link.aps.org/doi/10.1103/PhysRevC.108.045803}

\bibitem{PhysRevC.109.065807}
B.~Gao, Y.~Yan, M.~Harada,
  \href{https://link.aps.org/doi/10.1103/PhysRevC.109.065807}{Reconciling
  constraints from the supernova remnant {HESS} {J}1731-347 with the parity
  doublet model}, Phys. Rev. C 109 (2024) 065807.
\newblock \href {https://doi.org/10.1103/PhysRevC.109.065807}
  {\path{doi:10.1103/PhysRevC.109.065807}}.
\newline\urlprefix\url{https://link.aps.org/doi/10.1103/PhysRevC.109.065807}

\bibitem{LI2023138062}
J.~J. Li, A.~Sedrakian,
  \href{https://www.sciencedirect.com/science/article/pii/S0370269323003969}{Baryonic
  models of ultra-low-mass compact stars for the central compact object in hess
  j1731-347}, Phys. Lett. B 844 (2023) 138062.
\newblock \href
  {https://doi.org/https://doi.org/10.1016/j.physletb.2023.138062}
  {\path{doi:https://doi.org/10.1016/j.physletb.2023.138062}}.
\newline\urlprefix\url{https://www.sciencedirect.com/science/article/pii/S0370269323003969}

\bibitem{PhysRevC.108.025806}
L.~Brodie, A.~Haber,
  \href{https://link.aps.org/doi/10.1103/PhysRevC.108.025806}{Nuclear and
  hybrid equations of state in light of the low-mass compact star in {HESS}
  {J}1731-347}, Phys. Rev. C 108 (2023) 025806.
\newblock \href {https://doi.org/10.1103/PhysRevC.108.025806}
  {\path{doi:10.1103/PhysRevC.108.025806}}.
\newline\urlprefix\url{https://link.aps.org/doi/10.1103/PhysRevC.108.025806}

\bibitem{PhysRevD.110.043026}
M.~Mariani, I.~F. Ranea-Sandoval, G.~Lugones, M.~G. Orsaria,
  \href{https://link.aps.org/doi/10.1103/PhysRevD.110.043026}{Could a slow
  stable hybrid star explain the central compact object in {HESS}
  {J}1731-347?}, Phys. Rev. D 110 (2024) 043026.
\newblock \href {https://doi.org/10.1103/PhysRevD.110.043026}
  {\path{doi:10.1103/PhysRevD.110.043026}}.
\newline\urlprefix\url{https://link.aps.org/doi/10.1103/PhysRevD.110.043026}

\bibitem{Li_2024}
J.~J. Li, A.~Sedrakian, M.~Alford,
  \href{https://dx.doi.org/10.3847/1538-4357/ad4295}{Hybrid {S}tar {M}odels in
  the {L}ight of {N}ew {M}ultimessenger {D}ata}, Astrophys. J. 967~(2) (2024)
  116.
\newblock \href {https://doi.org/10.3847/1538-4357/ad4295}
  {\path{doi:10.3847/1538-4357/ad4295}}.
\newline\urlprefix\url{https://dx.doi.org/10.3847/1538-4357/ad4295}

\bibitem{Sagun_2023}
V.~Sagun, E.~Giangrandi, T.~Dietrich, O.~Ivanytskyi, R.~Negreiros,
  C.~Providência, \href{https://dx.doi.org/10.3847/1538-4357/acfc9e}{What {I}s
  the {N}ature of the {HESS} {J}1731-347 {C}ompact {O}bject?}, Astrophys. J.
  958~(1) (2023) 49.
\newblock \href {https://doi.org/10.3847/1538-4357/acfc9e}
  {\path{doi:10.3847/1538-4357/acfc9e}}.
\newline\urlprefix\url{https://dx.doi.org/10.3847/1538-4357/acfc9e}

\bibitem{Koliogiannis-2020}
P.~S. Koliogiannis, C.~C. Moustakidis,
  \href{https://link.aps.org/doi/10.1103/PhysRevC.101.015805}{Effects of the
  equation of state on the bulk properties of maximally rotating neutron
  stars}, Phys. Rev. C 101 (2020) 015805.
\newblock \href {https://doi.org/10.1103/PhysRevC.101.015805}
  {\path{doi:10.1103/PhysRevC.101.015805}}.
\newline\urlprefix\url{https://link.aps.org/doi/10.1103/PhysRevC.101.015805}

\bibitem{Kaplan-1988}
D.~Kaplan, A.~Nelson, Kaon condensation in dense matter., Nuc. Phys. A 479
  (1988) 273--284.
\newblock \href {https://doi.org/https://doi.org/10.1016/0375-9474(88)90442-3}
  {\path{doi:https://doi.org/10.1016/0375-9474(88)90442-3}}.

\bibitem{Serot-1997}
B.~D. Serot, J.~D. Walecka,
  \href{https://doi.org/10.1142/S0218301397000299}{Recent {P}rogress in
  {Q}uantum {H}adrodynamics}, Int. J. Mod. Phys. E 06~(04) (1997) 515--631.
\newblock \href
  {http://arxiv.org/abs/https://doi.org/10.1142/S0218301397000299}
  {\path{arXiv:https://doi.org/10.1142/S0218301397000299}}, \href
  {https://doi.org/10.1142/S0218301397000299}
  {\path{doi:10.1142/S0218301397000299}}.
\newline\urlprefix\url{https://doi.org/10.1142/S0218301397000299}

\bibitem{Moustakidis-2007}
V.~P. Psonis, C.~C. Moustakidis, S.~E. Massen,
  \href{https://doi.org/10.1142/S0217732307023572}{Nuclear symmerty energy
  effects on neutron stars properties}, Mod. Phys. Lett. A 22~(17) (2007)
  1233--1253.
\newblock \href {https://doi.org/10.1142/S0217732307023572}
  {\path{doi:10.1142/S0217732307023572}}.
\newline\urlprefix\url{https://doi.org/10.1142/S0217732307023572}

\bibitem{Shapiro-1983}
S.~Shapiro, S.~Teukolsky, Black {H}oles, {W}hite {D}warfs, and {N}eutron Stars,
  John Wiley and Sons, New York, 1983.

\bibitem{Glendenning-2000}
N.~Glendenning, Compact {S}tars: {N}uclear {P}hysics, {P}article {P}hysics, and
  {G}eneral {R}elativity, Springer, Berlin, 2000.

\bibitem{Flanagan-08}
E.~E. Flanagan, T.~Hinderer,
  \href{https://link.aps.org/doi/10.1103/PhysRevD.77.021502}{Constraining
  neutron-star tidal love numbers with gravitational-wave detectors}, Phys.
  Rev. D 77 (2008) 021502.
\newblock \href {https://doi.org/10.1103/PhysRevD.77.021502}
  {\path{doi:10.1103/PhysRevD.77.021502}}.
\newline\urlprefix\url{https://link.aps.org/doi/10.1103/PhysRevD.77.021502}

\bibitem{Hinderer-08}
T.~Hinderer, \href{https://doi.org/10.1086/533487}{Tidal {L}ove {N}umbers of
  {N}eutron {S}tars}, Astrophys. J. 677~(2) (2008) 1216--1220.
\newblock \href {https://doi.org/10.1086/533487} {\path{doi:10.1086/533487}}.
\newline\urlprefix\url{https://doi.org/10.1086/533487}

\bibitem{Baym-71}
G.~{Baym}, C.~{Pethick}, P.~{Sutherland}, {{T}he {G}round {S}tate of {M}atter
  at {H}igh {D}ensities: {E}quation of {S}tate and {S}tellar {M}odels},
  Astrophys. J. 170 (1971) 299.
\newblock \href {https://doi.org/10.1086/151216} {\path{doi:10.1086/151216}}.

\bibitem{Kanakis-Petousis-2024}
A.~Kanakis-Pegios, V.~Petousis, M.~Veselsk\'y, J.~Leja, C.~C. Moustakidis,
  \href{https://link.aps.org/doi/10.1103/PhysRevD.109.043028}{Constraints for
  the {X}17 boson from compact objects observations}, Phys. Rev. D 109 (2024)
  043028.
\newblock \href {https://doi.org/10.1103/PhysRevD.109.043028}
  {\path{doi:10.1103/PhysRevD.109.043028}}.
\newline\urlprefix\url{https://link.aps.org/doi/10.1103/PhysRevD.109.043028}

\bibitem{Arzoumanian-2018}
Z.~Arzoumanian, et~al., \href{https://dx.doi.org/10.3847/1538-4365/aab5b0}{The
  {NANOG}rav 11-year {D}ata {S}et: {H}igh-precision {T}iming of 45
  {M}illisecond {P}ulsars}, Astrophys. J. Suppl. S. 235~(2) (2018) 37.
\newblock \href {https://doi.org/10.3847/1538-4365/aab5b0}
  {\path{doi:10.3847/1538-4365/aab5b0}}.
\newline\urlprefix\url{https://dx.doi.org/10.3847/1538-4365/aab5b0}

\bibitem{Antoniadis-2013}
J.~Antoniadis, et~al.,
  \href{https://www.science.org/doi/abs/10.1126/science.1233232}{A {M}assive
  {P}ulsar in a {C}ompact {R}elativistic {B}inary}, Sci. 340~(6131) (2013)
  1233232.
\newblock \href
  {http://arxiv.org/abs/https://www.science.org/doi/pdf/10.1126/science.1233232}
  {\path{arXiv:https://www.science.org/doi/pdf/10.1126/science.1233232}}, \href
  {https://doi.org/10.1126/science.1233232}
  {\path{doi:10.1126/science.1233232}}.
\newline\urlprefix\url{https://www.science.org/doi/abs/10.1126/science.1233232}

\bibitem{Cromartie-2020}
H.~Cromartie, et~al.,
  \href{https://doi.org/10.1038/s41550-019-0880-2}{Relativistic {S}hapiro delay
  measurements of an extremely massive millisecond pulsar.}, Nat. Astron. 4
  (2020) 72--76.
\newblock \href {https://doi.org/https://doi.org/10.1038/s41550-019-0880-2}
  {\path{doi:https://doi.org/10.1038/s41550-019-0880-2}}.
\newline\urlprefix\url{https://doi.org/10.1038/s41550-019-0880-2}

\bibitem{Romani-2022}
R.~W. Romani, D.~Kandel, A.~V. Filippenko, T.~G. Brink, W.~Zheng,
  \href{https://dx.doi.org/10.3847/2041-8213/ac8007}{{PSR J0952-0607}: {T}he
  {F}astest and {H}eaviest {K}nown {G}alactic {N}eutron {S}tar}, Astrophys. J.
  Lett. 934~(2) (2022) L17.
\newblock \href {https://doi.org/10.3847/2041-8213/ac8007}
  {\path{doi:10.3847/2041-8213/ac8007}}.
\newline\urlprefix\url{https://dx.doi.org/10.3847/2041-8213/ac8007}

\bibitem{Abbott-2019}
B.~Abbott, et~al., Properties of the {B}inary {N}eutron {S}tar {M}erger
  {GW170817}., Phys. Rev. X 9 (2019) 011001.
\newblock \href {https://doi.org/https://doi.org/10.1103/PhysRevX.9.011001}
  {\path{doi:https://doi.org/10.1103/PhysRevX.9.011001}}.

\bibitem{Koliogiannis-2021}
P.~S. Koliogiannis, C.~C. Moustakidis,
  \href{https://dx.doi.org/10.3847/1538-4357/abe542}{Thermodynamical
  {D}escription of {H}ot, {R}apidly {R}otating {N}eutron {S}tars,
  {P}rotoneutron {S}tars, and {N}eutron {S}tar {M}erger {R}emnants}, Astrophys.
  J. 912~(1) (2021) 69.
\newblock \href {https://doi.org/10.3847/1538-4357/abe542}
  {\path{doi:10.3847/1538-4357/abe542}}.
\newline\urlprefix\url{https://dx.doi.org/10.3847/1538-4357/abe542}

\bibitem{Doroshenko-2022}
V.~Doroshenko, V.~Suleimanov, G.~Pühlhofer, A.~Santangelo,
  \href{https://doi.org/10.1038/s41550-022-01800-1}{A strangely light neutron
  star within a supernova remnant.}, Nat. Astron. 6 (2022) 1444–1451.
\newblock \href {https://doi.org/https://doi.org/10.1038/s41550-022-01800-1}
  {\path{doi:https://doi.org/10.1038/s41550-022-01800-1}}.
\newline\urlprefix\url{https://doi.org/10.1038/s41550-022-01800-1}

\bibitem{10.1093/mnras/stae2398}
Y.~Kini, et~al., \href{https://doi.org/10.1093/mnras/stae2398}{Constraining the
  properties of the thermonuclear burst oscillation source {XTE} {J}1814-338
  through pulse profile modelling}, Mon. Not. R. Astron. Soc. 535~(2) (2024)
  1507--1525.
\newblock \href
  {http://arxiv.org/abs/https://academic.oup.com/mnras/article-pdf/535/2/1507/60628222/stae2398.pdf}
  {\path{arXiv:https://academic.oup.com/mnras/article-pdf/535/2/1507/60628222/stae2398.pdf}},
  \href {https://doi.org/10.1093/mnras/stae2398}
  {\path{doi:10.1093/mnras/stae2398}}.
\newline\urlprefix\url{https://doi.org/10.1093/mnras/stae2398}

\bibitem{Salmi_2024}
T.~Salmi, et~al., \href{https://dx.doi.org/10.3847/1538-4357/ad81d2}{A {NICER}
  {V}iew of {PSR} {J}1231-1411: {A} {C}omplex {C}ase}, Astrophys. J. 976~(1)
  (2024) 58.
\newblock \href {https://doi.org/10.3847/1538-4357/ad81d2}
  {\path{doi:10.3847/1538-4357/ad81d2}}.
\newline\urlprefix\url{https://dx.doi.org/10.3847/1538-4357/ad81d2}

\bibitem{KaonReview}
F.~D. Clemente, M.~Casolino, A.~Drago, M.~Lattanzi, C.~Ratti,
  \href{https://arxiv.org/abs/2404.12094}{Strange quark matter as dark matter:
  40 years later, a reappraisal} (2024).
\newblock \href {http://arxiv.org/abs/2404.12094} {\path{arXiv:2404.12094}}.
\newline\urlprefix\url{https://arxiv.org/abs/2404.12094}

\end{thebibliography}
\end{document}